\documentclass[aps,pre,preprint,groupedaddress,showpacs]{revtex4-1}
\usepackage{graphicx,amsmath}
\begin{document}

\title{Diffusion equations from kinetic models with non-conserved momentum}


\author{P.L. Garrido}
\email[]{garrido@onsager.ugr.es}
\affiliation{Instituto Carlos I de F{\'\i}sica Te{\'o}rica y Computacional. Universidad de Granada. E-18071 Granada. Spain }

\author{Joel L. Lebowitz}
\email[]{lebowitz@math.rutgers.edu}
\affiliation{Dept. of Mathematics and Physics, Rutgers University, Piscataway, New Jersey 08854, USA }

\date{\today}
\begin{abstract}
We derive diffusive macroscopic equations for the particle and energy density of a system whose time evolution is described by a kinetic equation for the one particle position and velocity function $f(r,v,t)$ that consists of a part that conserves energy and momentum such as the Boltzmann equation and an external randomization of the particle velocity {\it directions} that breaks the momentum conservation. Rescaling space and time by $\epsilon$ and $\epsilon^2$ respectively and carrying out a Hilbert expansion in $\epsilon$ around a local equilibrium Maxwellian yields coupled diffusion equations with specified Onsager coefficients for the particle and energy density. Our analysis includes a system of hard disks at intermediate densities by using the Enskog equation for the collision kernel.  
\end{abstract}
\pacs{18-3e}
\maketitle

\section{Introduction}
The derivation of macroscopic equations such as the Navier-Stokes equations or the diffusion equation from the microscopic Hamiltonian dynamics governing the motion of the atomic constituents of matter is one of the central problems of nonequilibrium statistical mechanics \cite{Leb}.This is a very difficult problem whose full solution is currently out of sight. An intermediate step is the derivation of macroscopic equations from kinetic equations, such as the Boltzmann equation for gases, which describe the time evolution of the one particle distribution function $f(r,v,t)$ in the one particle position and velocity space, c.f.\cite{Cer}. These nonlinear equations have local conservation laws arising from the microscopic dynamics, corresponding to particle, momentum and (kinetic) energy densities. These conserved quantities evolve  on a suitable coarse grained space-time scales according to macroscopic equations. 
The actual mathematical derivation for all the conserved fields is a daunting task  which is far from complete even for this model mesoscopic evolution \cite{Esposito}.

In this note we study a model kinetic system in which there are only two conserved quantities: the particle and energy densities. We eliminate momentum conservation by considering situations in which the particles undergo, in addition to interparticle collisions, also collisions which do not conserve momentum, e.g. collisions with a substrate of randomly fixed scatterers. We then use diffusive space time scaling to derive coupled macroscopic diffusion equations for the particle and energy densities in two dimensions for different models of inter-particle collisions: BGK, Boltzmann and Enskog. The 3d case will be considered in a separate more mathematical paper \ref{EGLM}. We will use these 2d results to compare the transport coefficients obtained from Enskog equation with the results of molecular dynamic simulations of hard discs evolving according to Hamiltonian dynamics plus virtual collisions on the surface of a cylinder in which the top and bottom wall are kept at different temperatures \cite{Gar}. Fluctuations in the nonequilibrium stationary state of that system will be compared with the predictions of macroscopic fluctuation theory (MFT) \cite{Bertini}. 

\section{The general scheme}
Let $f(r,v,t)$ represent the density at time $t$ of particles at a position $r$ of a periodic torus $\Lambda$, $r\in \Lambda\subset {\rm I\!R}^d$, with a vector velocity $v\in {\rm I\!R}^d$. The distribution $f$ is assumed to satisfy an equation of the form,
\begin{equation}
\partial_t f+v\cdot\nabla f=Q_C(f)+\alpha Q_D(f)=Q(f) \label{BE}
\end{equation} 
where the nonlinear term $Q_C(f)$ represents the time variation of $f$ due to particle interactions e.g. the Boltzmann collision term\cite{Cer}, and the linear term $Q_D(f)$ represents collisions with a background. $Q_C(f)$ satisfies the basic conservation laws of the underlying microscopic dynamics, i.e. the density, momentum and the kinetic energy. This is expressed by the equation
\begin{equation}
\int dv\, \phi(v) Q_C(f)=0\quad\quad\forall f\label{inv0}
\end{equation}
where $\phi(v)=1$, $v$ and $v^2$ respectively. $Q_D(f)$ on the other hand only conserve the density and energy. 

Using the diffusive scaling: $\tau=\epsilon^2 t$ and $x=\epsilon r$, where the parameter $\epsilon$ is the ratio of some microscopic mean free path to a typical macroscopic length, the rescaled distribution reads \cite{Esposito}:
\begin{equation}
g(x,v,\tau)=f(\epsilon^{-1}r,v,\epsilon^{-2}\tau)\label{scal}
\end{equation}
and (\ref{BE}) takes the form:
\begin{equation}
\epsilon^2\partial_\tau g+\epsilon v\cdot\nabla g=Q_C(g)+\alpha Q_D(g) \label{BE2}
\end{equation}
We now make a formal $\epsilon$ expansion of $g$: 
\begin{equation}
g=g_0+\epsilon g_1+O(\epsilon^2)
\end{equation}
usually called {\it Hilbert expansion} \cite{Cer}. Substituting this into eq. (\ref{BE2}) we get order by order in $\epsilon$ the infinite set of equations:
\begin{eqnarray}
O(\epsilon^0):&&\quad Q(g_0)=0\label{e0}\\
O(\epsilon):&&\quad v\cdot\nabla g_0= Q_1(g_0,g_1)\label{e1}\\
O(\epsilon^2):&&\quad\partial_\tau g_0+ v\cdot\nabla g_1=Q_2(g_0,g_1,g_2)\label{e2}
\end{eqnarray}
where we have set $Q(g)=Q(g_0)+\epsilon Q_1(g_0,g_1)+\epsilon^2 Q_2(g_0,g_1,g_2)+O(\epsilon^3)$. The equation for the collision invariants, $\phi(v)=1$ and $v^2$, reads
\begin{equation}
\epsilon^2\partial_\tau \int dv\, \phi(v) g(x,v,\tau)+\epsilon\sum_i\partial_i\int dv\,\phi(v)v_i g(x,v,\tau)=0 \label{inv00}
\end{equation}
 and  its first nontrivial order ($\epsilon^2$) is:
\begin{equation}
\partial_\tau \int dv\, \phi(v) g_0(x,v,\tau)+\sum_i\partial_i\int dv\,\phi(v)v_i g_1(x,v,\tau)=0 \label{inv}
\end{equation} 
which is going to be the key equation for us. The solution of eq. (\ref{e0}) has the form:
\begin{equation}
g_0(x,v,\tau)=\bar g_0(v;\phi(x,\tau))
\end{equation}
with $\phi(x,\tau)=\int_ v dv\,\phi(v) g(x,v,\tau)$.
 That is, $g_0$ depends on $x$ and $\tau$ only through the fields $\phi(x,\tau)$. We will then look for a solution of  eq.(\ref{e1}) that depends on $x$ only through $\phi$ and its gradients: $g_1(x,v,\tau)=\bar g_1(v;[\phi(x,\tau)])$. Consequently the equation for the invariants (\ref{inv}) would be a set of closed kinetic equations for $\phi(x,\tau)$.
 Thus, given $Q$, we only need to get $g_0$, $g_1$ to build the set of diffusive equations through eq. (\ref{inv}) .

\section{The model}

Restricting ourselves to two dimensions we shall consider various forms of $Q_C(f)$: the Boltzmann (B) collision term, the Bhatnagar-Gross-Krook (BGK) kernel and the Enskog (E) modified B term.  The solution of $Q_C(g_{0})=0$ is the Maxwellian 
\begin{equation}
M(v;\bar n,\bar u,\bar T)=\frac{\bar n}{2\pi \bar T}\exp\left[-\frac{(v-\bar u)^2}{2\bar T}\right]\equiv \bar n m(v;\bar u,\bar T)\label{maxC2}
\end{equation}
 
The background collision term $Q_D(f)$ will have allways the form (in the rescaled variables):
\begin{equation}
Q_D(g)=\int\, d\hat n \left[g(x,v'',\tau)-g(x,v,\tau) \right]\label{QR}
\end{equation}
where $v''$ is the velocity after the randomization;
\begin{equation}
v''=v-2\hat n(v\cdot \hat n)
\end{equation}
The collision invariants for $Q_D$ are $\phi(v)=1$ and $v^2$ (it does not conserve momentum).

The solution of the equation $Q(g_0)=0$ is:
\begin{equation}
g_{0}(x,v,\tau)=M(v;n(x,\tau),0, T(x,\tau))\label{max}
\end{equation}
where 
\begin{eqnarray}
n(x,\tau)&=&\int dv\, g(x,v,\tau)\quad\quad\quad  \text{\it Particle density}\nonumber\\
u_i(x,\tau)&=&n(x,\tau)^{-1}\int dv\, v_i g(x,v,\tau)\quad (i=1,2)\quad \text{\it Hydrodynamic velocity}\label{def}\\
T(x,\tau)&=&n(x,\tau)^{-1}\int dv\, \frac{1}{2}(v-u)^2 g(x,v,\tau)\quad \text{\it Temperature}\nonumber
\end{eqnarray}
The relevant fields at the diffusive level are $n(x,\tau)$ and $T(x,\tau)$. Obviously, when $\alpha=0$, we will recover the $Q_C$ invariants. The transition from $\alpha$ nonzero to zero is singular.

\section{The BGK $Q_C$}

We first consider the BGK approximation \cite{BGK}. $Q_C$ is chosen to be a Maxwellian minus the one-particle distribution:
\begin{equation}
Q_C^{BGK}(g)=\nu \left[M(v;n(x,\tau),u(x,\tau),T(x,\tau))-g(x,v,\tau)\right]\label{BGK7}
\end{equation}
with $\nu$ a positive rate constant and $n(x,\tau)$, $u(x,\tau)$ and $T(x,\tau)$ are defined in eqs. (\ref{def}). Then $Q(g)=Q_C^{BGK}(g)+Q_D(g)$ and the zeroth term in the $\epsilon$ expansion, $g_0(x,v,\tau)$, is the Maxwellian \ref{maxC2} with zero $\bar u$. This implies that $u(x,v,\tau)$ is of order $\epsilon$: 
\begin{equation}
u(x,v,\tau)=\int dv\, v g(x,v,\tau)=\epsilon\int dv\, v g_1(x,v,\tau)+ O(\epsilon^2)\equiv \epsilon u_1(x,v,\tau)+O(\epsilon^2) \label{udef}
\end{equation}
Therefore, we should include the  expansion of $u$ in $\epsilon$ in the maxwellian term of $Q_C^{BGK}$. 

The $\epsilon$ expansion of $Q$ then gives the terms:
\begin{eqnarray}
Q(g_0)&=&\nu\left[M(v;n(x,\tau),0,T(x,\tau))\right]-g_0(x,v,\tau)+\alpha\int dv\,\left[g_0(x,v'',\tau)-g_0(x,v,\tau) \right]\nonumber\\
Q_1(g_0,g_1)&=&\nu\left[ M(v;n(x,\tau),0,T(x,\tau)) \frac{v u_1(x,\tau)}{T(x,\tau)}-g_1(x,v,\tau)\right]\nonumber\\
&+&\alpha\int dv\,\left[g_1(x,v'',\tau)-g_1(x,v,\tau) \right]\nonumber
\end{eqnarray}
We substitute these expressions into eqs. (\ref{e0}) and (\ref{e1}) to get the first two terms of the $\epsilon$ expansion of $g$.
The first of these equations give us the expected result: $g_{0}(x,v,\tau)=M(v;n(x,\tau),0, T(x,\tau))$. From the second we get the value of $g_1$. We look for solutions of the form:
\begin{equation}
g_1(x,v,\tau)=M(v;n(x,\tau),0, T(x,\tau)) v\cdot A(w)\label{g1}
\end{equation}
where $A(w)$ is  an unknown vector that is a function of the modulus of $v$: $w=\vert v\vert$. Substituting (\ref{g1}) into eq. (\ref{e1}) and using the fact that $A(w'')=A(w)$ we get:
\begin{equation}
A(w)=-\frac{1}{\nu+2\pi\alpha}\left[\frac{\nu u_1}{T}-\frac{\nabla n}{n}-\left(\frac{w^2}{2T}-1\right)\frac{\nabla T}{T} \right]\label{A0}
\end{equation}
Observe that $g_1$ is a function of $u_1$. We substitute eq. (\ref{A0}) and eq. (\ref{g1}) into eq. (\ref{udef}) to find the value of $u_1$. In fact  we just need to know that $\langle v_i^2\rangle_0=nT$ and $\langle v_i^2 v^2\rangle_0=4nT^2$ where $\langle\cdot\rangle_0=\int dv\, \cdot g_0$. Finally the value of $u_1$ is,
\begin{equation}
u_1(x,v,\tau)=-\frac{\nabla(nT)}{2\pi\alpha n}\label{u1}
\end{equation}
Substituting (\ref{u1}) into eq. (\ref{A0}) we get
\begin{equation}
A(w)=-\frac{1}{2\pi\alpha}\left[\frac{\nabla n}{n}+\frac{1}{\nu+2\pi\alpha}\left(\nu+2\pi\alpha\left(\frac{w^2}{2T}-1\right)\right)\frac{\nabla T}{T} \right]\label{A}
\end{equation}
We can now use eq. (\ref{inv}) to obtain equations for $n$ and $T$:
\begin{eqnarray}
\partial_\tau n(x,\tau)&+&\nabla J_n(x,\tau)=0\nonumber\\
\partial_\tau(n(x,\tau)T(x,\tau))&+&\nabla J_h(x,\tau)=0\label{dif}
\end{eqnarray}
with
\begin{eqnarray}
J_n(x,\tau)&=&-\frac{1}{2\pi\alpha}\left[T\nabla n+n\nabla T \right]=-\frac{1}{2\pi\alpha}\nabla P\label{Jn}\\
J_h(x,\tau)&=&-\frac{T}{\pi\alpha}\left[T\nabla n+n\nabla T \left(1+\frac{2\pi\alpha}{\nu+2\pi\alpha}\right) \right]=-\frac{T}{\pi\alpha}\nabla P-\frac{2P}{\nu+2\pi\alpha}\nabla T\label{Jq}
\end{eqnarray}
where we have used $\langle v_i^2 v^4\rangle_0=24nT^3$ and $P(x,\tau)=n(x,\tau)T(x,\tau)$ for the pressure. Thus when and only when $J_n=0$ (or equivalently $u_1(x,\tau)=0$), the pressure is a global constant,  $P(x,\tau)=\bar P$. This will happens at equilibrium when $T$ and $n$ are independent of $x$ and also in the stationary state corresponding to a temperature gradient imposed  by the boundaries of the box. 

From eqs. (\ref{Jn}) and (\ref{Jq}) we can get the diffusion constant $D$ and the thermal conductivity $\kappa$. They are respectively:
\begin{equation}
J_n=-D\nabla n\quad \text{when}\,\, T=cte\, , \quad\quad J_h=-\kappa\nabla T\quad\text{when}\,\, J_n=0
\end{equation}
In our case these are:
\begin{equation}
D=\frac{T}{2\pi\alpha}\quad\quad \kappa=\frac{2nT}{\nu+2\pi\alpha} \label{TC}
\end{equation}
Observe that $D$ is proportional to the temperature and that the thermal conductivity is constant when the pressure is constant. That is, for the stationary state in a box with a temperature gradient and $J_n=0$, the temperature profile will be linear for this form of $Q(f)$. 

Finally we are interested in the Onsager's coefficients \cite{Onsager} defined by
\begin{eqnarray}
J_n(x,\tau)&=&L_{nn}\nabla\frac{-\mu}{T}+L_{nh}\nabla\frac{1}{T}\label{Ln}\\
J_h(x,\tau)&=&L_{hn}\nabla\frac{-\mu}{T}+L_{hh}\nabla\frac{1}{T}\label{Lq}
\end{eqnarray}
where $\mu$ is the chemical potential for the underlying local equilibrium state. In our case $\mu$ is the one corresponding to an ideal gas:
\begin{equation}
\mu=T\log\frac{n}{T}+C T
\end{equation}
where $C$ is a constant. Forming the gradient of $\mu/T$ and comparing terms with eqs. (\ref{Jn}) and (\ref{Jq}) we find:
\begin{equation}
L_{nn}=\frac{nT}{2\pi\alpha}\quad ;\quad L_{nh}=L_{hn}=\frac{nT^2}{\pi\alpha}\quad ;\quad L_{hh}=\frac{2nT^3}{\pi\alpha}\frac{\nu+3\pi\alpha}{\nu+2\pi\alpha}\label{Ons}
\end{equation}

In the Appendix we derive the 	kinetic equations and the Onsager coefficients for the  $\alpha=0$ case in the BGK approximation. In particular, for the non-convective case (u=0) we find:
\begin{equation}
J_n=-\frac{\nabla(nT)}{\nu}\quad\quad J_h=-\frac{2\nabla(nT^2)}{\nu}
\end{equation}
that implies
\begin{equation}
D=\frac{T}{\nu}\quad\quad \kappa=\frac{2nT}{\nu}
\end{equation}

\section{The Boltzmann collision kernel}

We can now follow the strategy used in the BGK approximation to get the diffusion equations for the Boltzmann $Q_C$ for hard disc collisions: 
\begin{equation}
Q_C(g)=\frac{b}{2}\int d\hat n\int dv_2\,\vert (v-v_2)\cdot \hat n \vert\left[ g(x,v_2',\tau)g(x,v',\tau)-g(x,v_2,\tau)g(x,v,\tau)\right] \label{QB}
\end{equation}
where $b$ is the cross section, equal to the hard disk diameter, $\hat n=(\cos\psi,\sin\psi)$ is a unit vector and the integral extends over $\psi\in[0,2\pi]$, $v$ is the velocity of the reference particle distribution and $v_2$ is the velocity of the particle colliding with the reference one, and
 $v'$ and $v_2'$ are the system velocities after the collision:
\begin{eqnarray}
v'&=&v-\hat n ((v-v_2)\cdot \hat n)\nonumber\\
v_2'&=&v_2+\hat n((v-v_2)\cdot \hat n)\nonumber
\end{eqnarray}

The solution $g_0$ of eq. (\ref{e0}) is still $g_0(x,v,\tau)=M(n(x,\tau),0,T(x,\tau))$. The solution of eq. (\ref{e1}) for $g_1$ is discused in Appendix III where we show that the currents are:
\begin{eqnarray}
J_n&=&-\frac{1}{2\pi\alpha}\nabla P\nonumber \\
J_h&=&-\frac{T}{\pi\alpha}\nabla P+\left[ \frac{P}{2\pi\alpha}+\frac{n}{2}\bar a_{22}^{(\tilde N)}\right]\nabla T
\end{eqnarray}
where $P=nT$ and $\bar a_{22}^{(\tilde N)}$ is a coefficient that depends on the truncation of a known infinite matrix into a $\tilde N$ dimensional one.

The diffusion coefficient and the Onsager's coefficients $L_{nn}$ and $L_{nh}$ are the same as in the BGK case:
\begin{eqnarray}
D&=&\frac{T}{2\pi\alpha}\nonumber\\
L_{nn}&=&\frac{P}{2\pi\alpha}\nonumber\\
L_{nh}&=&L_{hn}=\frac{nT^2}{\pi\alpha}
\end{eqnarray}

$\kappa$ and $L_{hh}$  depend on $\tilde N$. We can get an explicit expression for them for some particular $\tilde N$ values:

\begin{itemize}
\item $\tilde N=2$:
\begin{equation}
 L_{hh}^{(2)}=\frac{2nT^3}{\pi\alpha}\left[1+\frac{\pi\alpha}{2\pi\alpha+bn(\pi T)^{1/2}}\right]
\end{equation}
\begin{equation}
\kappa^{(2)}=\frac{2nT}{2\pi\alpha+bn(\pi T)^{1/2}}\label{Q2}
\end{equation}
\item $\tilde N=3$: 
\begin{equation}
L_{hh}^{(3)}=\frac{2nT^3}{\pi\alpha}\left[1+\frac{\alpha}{2}\frac{48\pi\alpha+39bnT^{1/2}}{48\pi\alpha^2+63\pi^{1/2}b\alpha n T+19b^2 n^2 T}\right]
\end{equation}
\begin{equation}
\kappa^{(3)}=\frac{3nT}{\pi}\frac{16\pi\alpha+13 b n (\pi T)^{1/2}}{48\pi\alpha^2+63 b\alpha n (\pi T)^{1/2}+19 b^2 n^2 T}
\end{equation}
\end{itemize}
Other values of $\tilde N$ can be studied but the algebraic expressions are much larger and no relevant information arise from them. Below we will compare numerically all the different methods and several values of $\tilde N$.

\section{The Enskog equation for intermediate densities}
The Boltzmann $Q_C$ is only valid in the very low density regime. In fact, the local equation of state is the ideal gas one. The Enskog equation tries to go beyond this limit by introducing a spatial dependence in the collision kernel $Q_C$ \cite{CE}:
\begin{eqnarray}
Q_C^{E}&=&b\int d\hat n\int_{(v_2-v)\cdot\hat n>0} dv_2\,\left((v_2-v)\cdot\hat n\right)\biggl[\chi(x+\frac{b}{2}\hat n)f(x+b\hat n,v_2',t)f(x,v',t)\nonumber\\
&-&\chi(x-\frac{b}{2}\hat n)f(x,v,t)f(x-b\hat n,v_2,t) \biggr]\label{CE1}
\end{eqnarray} 
where $\chi(x)$ is proportional to the equilibrium pair correlation function of hard disks at contact at the local density \cite{vB,Res,Gol}. 

As before, we introduce the diffusive scaling : $\tau=\epsilon^2 t$, $x=\epsilon r$. The scaled equation using (\ref{scal}) now reads:
\begin{eqnarray}
\epsilon^2\partial_\tau g+\epsilon v\cdot\nabla g=Q_{C,0}^{E}+\epsilon Q_{C,1}^{E}+\epsilon^2 Q_{C,2}^{E}+\alpha Q_D+O(\epsilon^3) \label{CE2}
\end{eqnarray} 
where
\begin{eqnarray}
Q_{C,0}^{E}&=&b\chi(r)\int d\hat n\int_{(v_2-v)\cdot\hat n>0} dv_2\,\left((v_2-v)\cdot\hat n\right)\biggl[g(r,v_2',\tau)g(r,v',\tau)
-g(r,v,\tau)g(r,v_2,\tau) \biggr]\nonumber\\
Q_{C,1}^{E}&=&-\frac{b^2}{2}\int d\hat n\int_{(v_2-v)\cdot\hat n>0} dv_2\,\left((v_2-v)\cdot\hat n\right)\biggl[(g(r,v_2',\tau)g(r,v',\tau)\nonumber\\
&+&g(r,v,\tau)g(r,v_2,\tau))\left(\hat n\cdot\nabla\chi\right)+2\chi\bigl(g(r,v',\tau)\left(\hat n\cdot\nabla g(r,v_2',\tau)\right)\nonumber\\
&+&g(r,v,\tau)\left(\hat n\cdot\nabla g(r,v_2,\tau)\right)\bigr)
 \biggr]\nonumber\\
 Q_{C,2}^{E}&=&\frac{b^3}{4}\int d\hat n\int_{(v_2-v)\cdot\hat n>0} dv_2\,\left((v_2-v)\cdot\hat n\right)\sum_{i,j}\hat n_i\hat n_j\biggl[
 2\chi\bigl(\partial_{ij}g(r,v_2',\tau)g(r,v',\tau)\nonumber\\
&-&\partial_{ij}g(r,v_2,\tau)g(r,v,\tau)\bigr)+2\partial_i\chi\bigl(\partial_j g(r,v_2',\tau)g(r,v',\tau)-\partial_j g(r,v_2,\tau)g(r,v,\tau)\bigr)\nonumber\\
&+&\frac{1}{2}\partial_{ij}\chi\left(g(r,v_2',\tau)g(r,v',\tau)
-g(r,v,\tau)g(r,v_2,\tau)\right)\biggr]\label{exp1}
\end{eqnarray}
where we have transformed $\chi(x)\rightarrow\chi(r)$. We can introduce now the Hilbert expansion $g=g_0+\epsilon g_1+\ldots$ and identifying $\epsilon$ orders at both sides of eq. (\ref{CE2}) we obtain:
\begin{eqnarray}
O(\epsilon^0)&:& Q_{C,0}^{E}(g_0)+Q_D(g_0)=0\label{CEe1}\\
O(\epsilon)&:&v\cdot\nabla g_0=Q_{C,0;1}^{E}(g_0,g_1)+Q_{D;1}(g_1)+Q_{C,1;0}^{E}(g_0)\label{CEe2}\\
O(\epsilon^2)&:&\partial_\tau g_0+v\cdot\nabla g_1=Q_{C,0;2}^{E}(g_0,g_1,g_2)+Q_{C,1;1}^{E}(g_0,g_1)\nonumber\\
&+&Q_{C,2;0}^{E}(g_0)+Q_{D;2}(g_2)\label{CEe3}
\end{eqnarray}
where
\begin{eqnarray}
Q_{C,i}(g)^{E}&=&Q_{C,i}^{E}(g_0+\epsilon g_1+\epsilon^2g_2+\ldots)\nonumber\\
&=&Q_{C,i;0}^{E}(g_0)+\epsilon Q_{C,i;1}^{E}(g_0,g_1)+\epsilon^2 Q_{C,i;2}^{E}(g_0,g_1,g_2)+O(\epsilon^3)\label{exp2}
\end{eqnarray}
We can solve the above equations for $g_0$, $g_1$, $...$ order by order. 
The general solution of eq. (\ref{CEe1}) is again the maxwellian: 
\begin{equation}
g_{0}(x,v,\tau)=M(v;n(x,\tau),0, T(x,\tau))
\end{equation}

The solution of eq. (\ref{CEe2}) for $g_1$ needs a bit more work and it is done explictly in Appendix IV. 

Once we know the solution $g_1$ we can use eq. (\ref{CEe3}) to get the diffusive equations for $n$ and $T$. We multiply both sides of eq. (\ref{CEe3}) by $\phi(v)$ ($=1,v^2$) and integrate with respect to $v$. Thus yields
\begin{eqnarray}
&&\partial_{\tau}\langle\phi(v)\rangle_0+\sum_{i}\langle v_i\phi(v)\Phi\rangle_0=\int dv\phi(v)\biggl[Q_{C,0;2}^{E}(g_0,g_1,g_2)\nonumber\\
&&+Q_{C,1;1}^{E}(g_0,g_1)
+Q_{C,2;0}^{E}(g_0)+Q_{D;2}(g_2) \biggr] \label{defe}
\end{eqnarray}
where we have used $g_1=g_0\Phi$ (see Appendix IV).
The $\phi(v)$ are invariants of $Q_{C,0}^{E}$ and $Q_D$ and therefore their integrals with $\phi(v)$ are equal to zero, order by order, in the $\epsilon$ expansion of $g$:
\begin{equation}
\int dv\phi(v)Q_{C,0;2}^{E}(g_0,g_1,g_2)=0\quad ,\quad \int dv\phi(v)Q_{D;2}(g_2) =0
\end{equation}
However, the remaining terms have not, a priori, the same collision invariants and they should be computed for each case.
\begin{itemize}
\item $\phi=1$: One can show by the time reversal symmetry of the equations that
\begin{equation}
\int dv\,Q_{C,1;1}^{E}(g_0,g_1)=0\quad , \quad \int dv\,Q_{C,2;0}^{E}(g_0)=0
\end{equation}
therefore
\begin{equation}
\partial_\tau n+\nabla J_n^{(\tilde N)}=0
\end{equation}
with
\begin{equation}
J_n^{(\tilde N)}=\frac{n}{2}f_1(n)\bar b_{11}\nabla n+\frac{n}{2}f_2(n)\bar b_{12}\frac{\nabla T}{T}\label{enkcurrd}
\end{equation}
where we have computed explicitely $\langle v_i\phi(v)\Phi\rangle_0$ and
\begin{equation}
\bar b_{ij}^{(\tilde N)}={\beta^{(i,\tilde N)}}^T F_{\tilde N}^{-1}\beta^{(j,\tilde N)}
\end{equation}
This notation was already explained in the Boltzmann kernel case. Observe that $\bar b_{11}^{(\tilde N)}$ and $\bar b_{12}^{(\tilde N)}$ do not depend on $\tilde N$ (see Appendix IV).

\item $\phi=v^2$: In this case we get
\begin{equation}
\langle v_i v^2\Phi\rangle_0=nTf_1(n)\left(\bar b_{12}^{(\tilde N)}+\frac{f_3(n)}{f_2(n)}\bar b_{11}^{(\tilde N)}\right)\partial_i n+n f_2(n)\left(\bar b_{22}^{(\tilde N)}+\frac{f_3(n)}{f_2(n)}\bar b_{12}^{(\tilde N)}\right)\partial_i T
\end{equation}
\begin{eqnarray}
\int dv\,v^2 Q_{C,1;1}^{E}(g_0,g_1)&=&
-\sum_i\partial_i\biggl[\frac{\pi b^2}{8}\chi n^2T\biggl(f_1(n)\left(3\bar b_{12}^{(\tilde N)}+(3\frac{f_3(n)}{f_2(n)}-2)\bar b_{11}^{(\tilde N)}\right)\partial_i n\nonumber\\ 
&+&f_2(n)\left(3\bar b_{22}^{(\tilde N)}+(3\frac{f_3(n)}{f_2(n)}-2)\bar b_{12}^{(\tilde N)}\right)\frac{\partial_i T}{T}
\biggr)
\biggr]
\end{eqnarray}
\begin{equation}
\int dv\,v^2Q_{C,2;0}^{E}(g_0)=\sum_i\partial_i\left[b^3\chi n^2(\pi T)^{1/2}\partial_i T\right]
\end{equation}
and the corresponding diffusion equation is then
\begin{equation}
\partial_{\tau}(nT)+\nabla J_h^{(\tilde N)}=0
\end{equation}
with
\begin{eqnarray}
J_h^{(\tilde N)}&=&\frac{1}{2}nTf_1(n)\left(\chi f_2(n)\bar b_{12}+\bar b_{11}\right)\nabla n\nonumber\\
&+&\frac{1}{2}n\left[f_2(n)\left(\chi f_2(n)\bar b_{22}^{(\tilde N)}+\bar b_{12}\right) -b^3\chi n(\pi T)^{1/2}\right]\nabla T
\end{eqnarray}
\end{itemize}

Before computing the diffusion coefficient, the thermal conductivity and the Onsager's coefficients we need to fix the value of $\chi(n)$. In the case of a hard disk system at equilibrium we know that its equation of state should be of the form
\begin{equation}
Q=T\rho(1+H(\rho))
\end{equation}
where $Q=P\pi b^2/4$, $P$ is the pressure and $\rho=\pi b^2 n/4$ is the areal density for a system of hard disks. While the exact form of $H(\rho)$ is not known there are good approximations to it for moderate densities \cite{Mulero}. The pair correlation function at contact is then
\begin{equation}
\chi(n)=\frac{H(\rho)}{2\rho}
\end{equation}

The local equilibrium equation of state for our system coincides with the equilibrium one for any value of the $\alpha$ parameter, i.e.
\begin{equation}
Q(x)=T(x)\rho(x)(1+\tilde H(\rho(x)))
\end{equation}
We know that at any stationary state with no local particle current, the local pressure should be constant all along the system.  

The absence of particle current at the stationary state implies that $J_n^{(\tilde N)}=0$ and from eq. (\ref{enkcurrd}):
\begin{equation}
f_1(n)\bar b_{11}\nabla n+f_2(n)\bar b_{12}\frac{\nabla T}{T}=0
\end{equation}

The constant pressure condition implies $\nabla Q=0$ and then, from the equation of state:
\begin{equation}
(1+\tilde H+\rho \tilde H')\nabla\rho+\rho(1+\tilde H)\frac{\nabla T}{T}=0
\end{equation}
both expressions should be correct for any $\nabla n$ and $\nabla T$ values which forces the relation:
\begin{equation}
\pi b^2\frac{1+\tilde H+\rho \tilde H'}{4\rho(1+\tilde H)}=\frac{f_1(n)\bar b_{11}}{f_2(n)\bar b_{12}}=\frac{f_1(n)}{2f_2(n)-f_3(n)}
\end{equation}
and we have used the explicit values of $\bar b_{11}$ and $\bar b_{12}$ obtained in Appendix IV. 
Observe that the $f$'s depend on $\chi(n)$ and therefore on $H(\rho)$. This equation relates $H(\rho)$ with $\tilde H(\rho)$. It is a matter of algebra to show that $\tilde H(\rho)=H(\rho)$ for all  values of $\alpha$.

The diffusion coefficient for any $\tilde N$ reads 
\begin{equation}
D=\frac{T}{2\pi\alpha\chi}\left[1+\pi b^2 n\chi+\frac{\pi}{2} b^2 n^2\chi'\right]
\end{equation}
and the thermal conductivity is now
\begin{equation}
\kappa^{(\tilde N)}=\frac{n\chi f_2(n)^2}{2\bar b_{11}}\left(\bar b_{12}^{2}-\bar b_{11}\bar b_{22}^{(\tilde N)}\right)+\frac{b^3}{2}n^2\chi(\pi T)^{1/2}
\end{equation}
In order to obtain the Onsager's coefficient we need first to compute the chemical potential of the local equilibrium reference system. 
We write the chemical potential as an $H$ dependent function using the fact that the free energy can be written as 
\begin{equation}
 f(\rho,T)=\log\frac{T}{\rho}-\int_{0}^{\rho}d\tilde\rho\frac{H(\tilde\rho)}{\tilde\rho}+cte
\end{equation}
Then, the chemical potential is:
\begin{equation}
\mu=-T\frac{\partial (\rho f)}{\partial \rho}=-T\left[\log\frac{T}{\rho}-\int_{0}^{\rho}d\tilde\rho\frac{H(\tilde\rho)}{\tilde\rho}-H(\rho)+cte\right]
\end{equation}
Thus, the terms $\nabla (-\mu/T)$ that appear on the definition of the Onsager's coefficients in eqs. (\ref{Ln}) and (\ref{Lq}) can be computed:
\begin{equation}
\nabla\left(\frac{-\mu}{T}\right)=\frac{\nabla T}{T}-(1+H(\rho)+\rho H(\rho)')\frac{\nabla n}{n}
\end{equation}
This yields:
\begin{eqnarray}
L_{nn}&=&\frac{nT}{2\pi\alpha\chi}\nonumber\\
L_{nh}&=&L_{hn}=\frac{nT^2}{\pi\alpha\chi}\left[1+\frac{\pi b^2}{4} n \chi \right]\nonumber\\
L_{hh}^{(\tilde N)}&=&\frac{nT^2}{2\chi}\left[\frac{T}{\pi\alpha}\left(3+\pi b^2n\chi\right)-\left(1+\frac{3}{8}\pi b^2n\chi\right)^2  \bar b_{22}^{(\tilde N)}+b^3\pi^{1/2} \chi^2 T^{1/2}\right]
\end{eqnarray}

In particular, for $\tilde N=2$ we get:
\begin{eqnarray}
\kappa^{(2)}&=&\frac{32 \pi ^{3/2} n^2 b^3 \sqrt{T} \chi  {\alpha }+n T \left(\pi  n b^2 \chi  \left((16+9 \pi ) n b^2 \chi +48\right)+64\right)}{32 \left(2\pi  {\alpha }+\sqrt{\pi } n b \sqrt{T} \chi \right)}\nonumber\\
L_{hh}^{(2)}&=&\frac{n T^{5/2}}{32 \pi  {\alpha } \left(2\sqrt{\pi }{\alpha}+n b \sqrt{T} \chi \right)}\times\nonumber\\ 
&&\biggl[32 \pi ^2 n b^3 \chi {\alpha }^2+\sqrt{\pi} \sqrt{T} {\alpha } \left(\pi  n b^2 \chi  \left((16+17 \pi ) n b^2 \chi +112\right)+192\right)\nonumber\\
&+&4 n b T \chi  \left(\pi  n b^2 \chi +4\right)^2\biggr]
\nonumber
\end{eqnarray}

\section{Results and Conclusions}
In order to compare the above results with numerical experiments it is convenient to define $\alpha$ as
\begin{equation}
\alpha=\frac{8\rho T^{1/2}}{\pi^{1/2}b}\frac{\tau_C}{\tau_D}
\end{equation}
where $\tau_C/\tau_D$ is the number of particle velocity randomizations done in a fixed time interval divided by the number of hard disk collisions that occur in such interval. From kinetic theory $\tau_C$ can be taken as the mean free time, $\tau_C=b\pi^{1/2}/8\rho T^{1/2}$ and $\tau_D=1/\alpha$. 
From now on we are going to do such substitution. In the BGK case we also are going to consider $\nu=1/2\tau_C$.

\begin{figure}[h!]
\begin{center}
\includegraphics[height=7cm,clip]{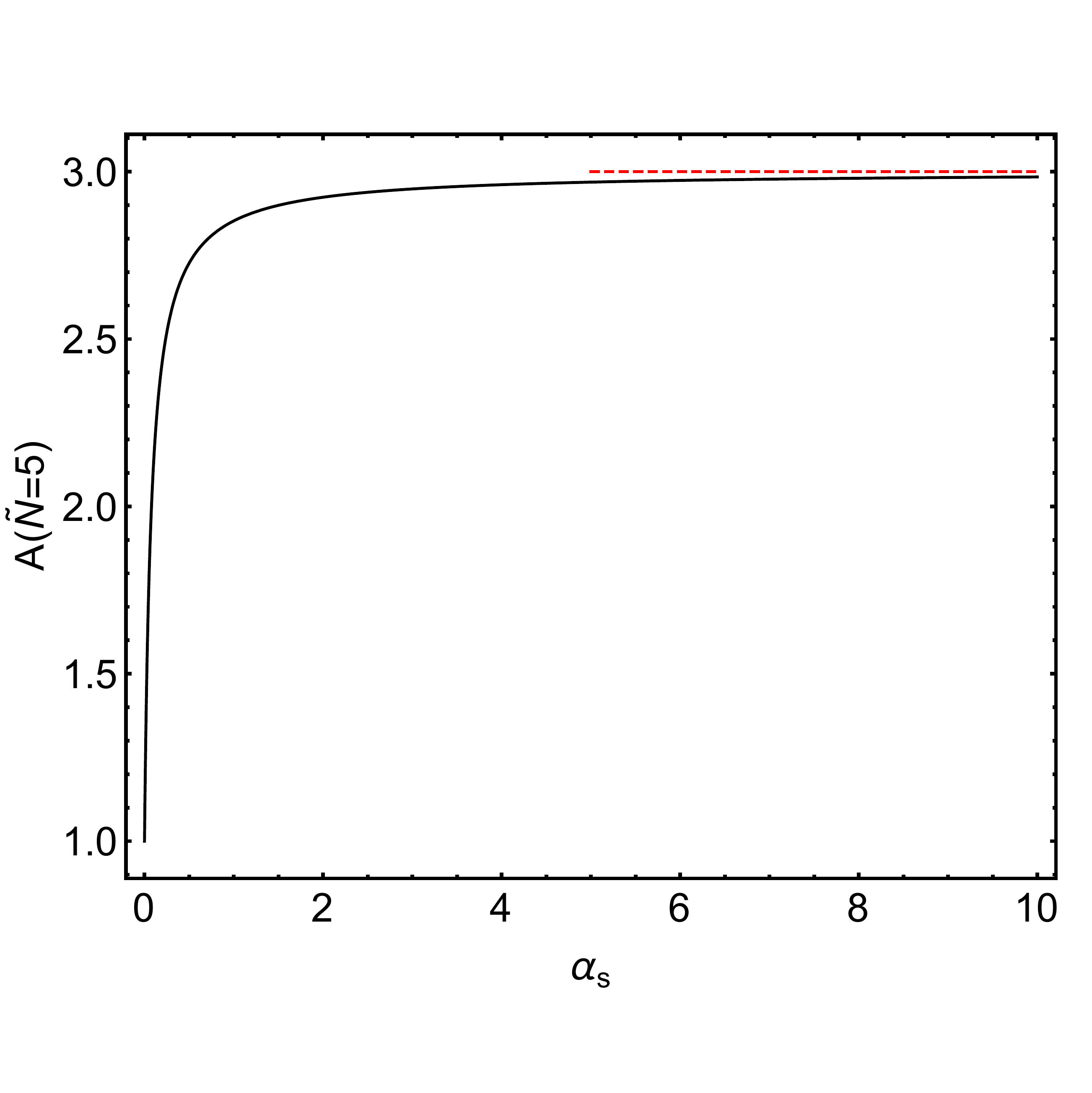}  %A_matrix_finite_size_2.nb
\includegraphics[height=7cm,clip]{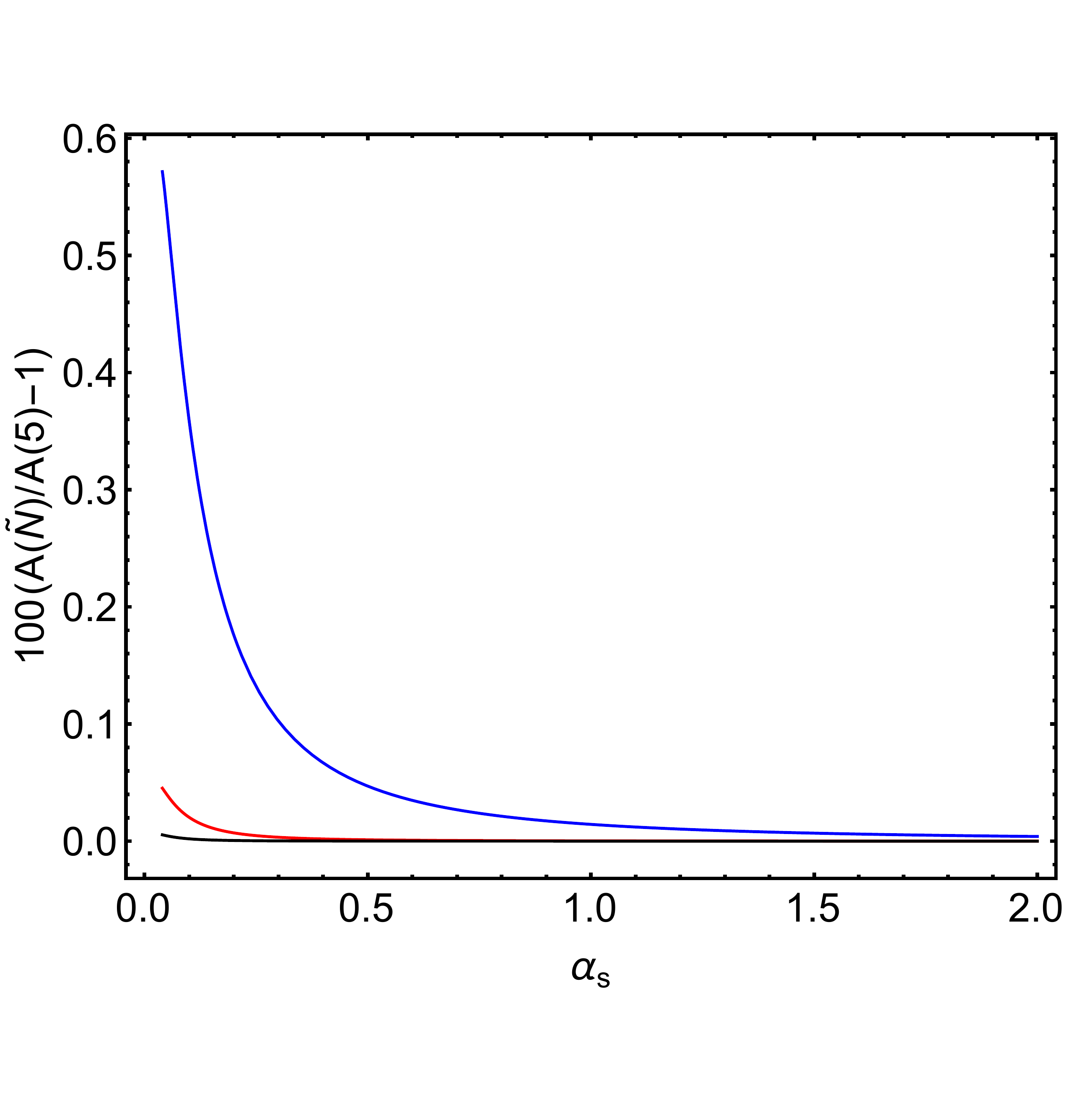}      
\end{center}
\kern -1cm
\caption{Left: $A(\tilde N=5)$ as a function of $\alpha_s$. The dashed red line is the asymptotic value of $A$ when $\alpha_s\rightarrow\infty$. Right: Relative error of $A(\tilde N)$ with respect $A(5)$ as a function of $\alpha_s$: Blue curve ($\tilde N=2$), Red curve ($\tilde N=3$) and Black curve ($\tilde N=4$).  \label{finite}}
\end{figure}

First we study the convergence with $\tilde N$ of $\bar a$'s. Let us remark a few properties of them:
\begin{equation}
\bar a_{11}^{(\tilde N)}=\bar a_{11}=-\frac{-b T^{1/2}}{8\pi^{1/2}\rho\alpha_s}\quad\forall\, \tilde N
\end{equation}
where $\alpha_s\equiv \tau_C/\tau_D$ and $A(\tilde N)=a_{22}^{(\tilde N)}/\bar a_{11}$ only depends on $\alpha_s$ and $\tilde N$. Also it can be seen that
\begin{equation}
\lim_{\alpha_s\rightarrow 0}A(\tilde N)=1\quad ,\quad \lim_{\alpha_s\rightarrow \infty}A(\tilde N)=3
\end{equation}
In figure \ref{finite} we show the behavior of $A(5)$ as a function of $\alpha_s$. We also show the relative error with respect $A(5)$. Observe how the worst case, $\tilde N=2$ and $\alpha_s\simeq 0$ the relative error is of $0.6\%$ and it decays uniformly with $\alpha_s$. Therefore we may conclude that the matrix truncation $F_{\tilde N}$ is a very good approximation of the real solution even for small values of $\tilde N$.

\begin{figure}[h!]
\begin{center}
\includegraphics[height=7cm,clip]{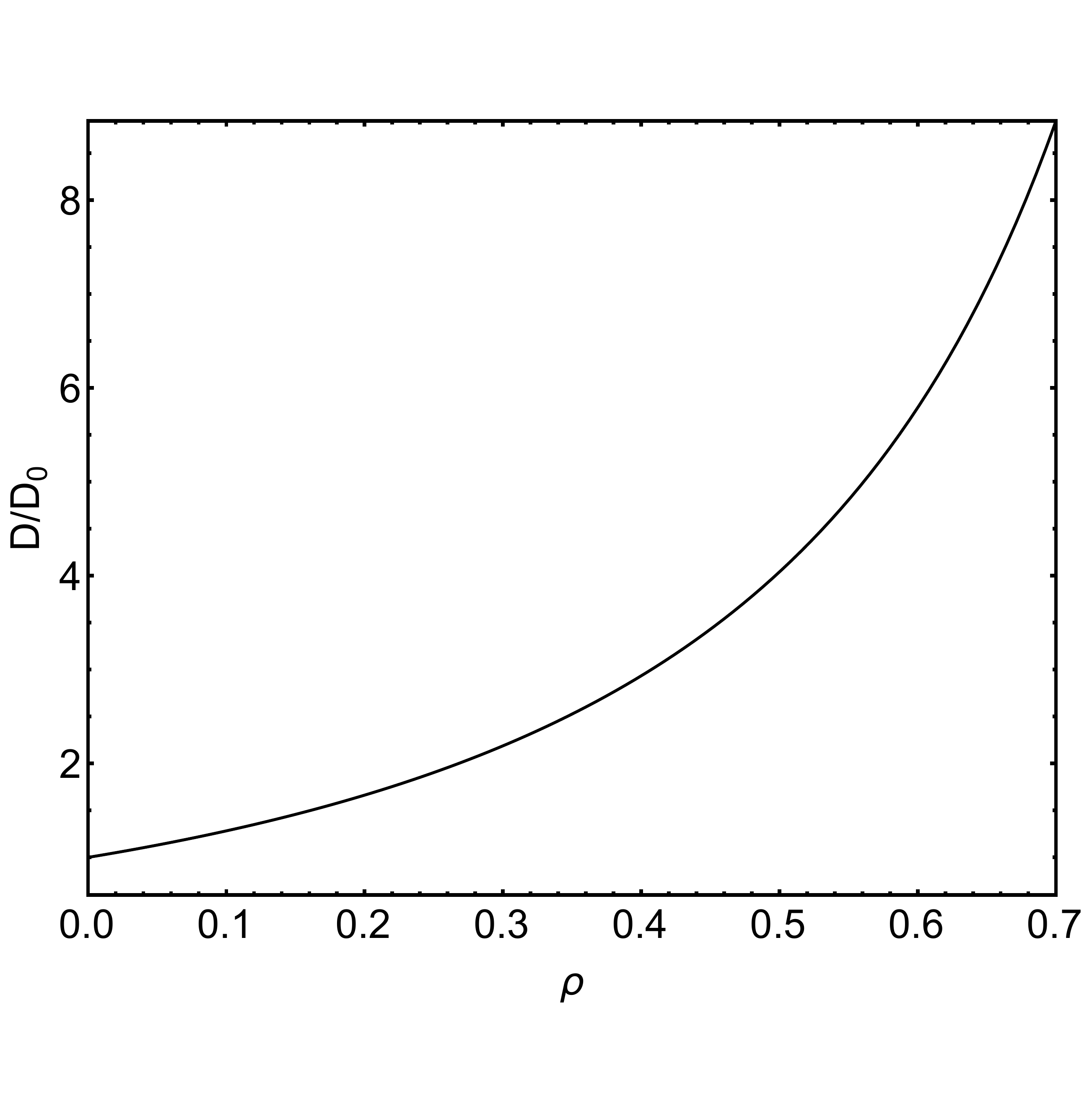}  %diffusion.nb
\end{center}
\kern -1cm
\caption{Diffusion coefficient for the Enskog equation vs areal density. $D_0$ is the diffusion coefficient for the BGK and Boltzmann cases. \label{diffusion}}
\end{figure}
The diffusion doefficient is given, in the three different approaches we have studied, by:
\begin{equation}
D_{BGK}=D_{B}=D_0\quad,D_{CE}=D_0\left[\chi^{-1}+4\rho+2\rho^2\chi^{-1}\chi'\right]
\end{equation}
where $D_0=bT^{1/2}/16\pi^{1/2}\alpha_s\rho$. See figure \ref{diffusion} for the behavior of $D/D_0$ as a function of $\rho$ for the Chapman-Enskog case and we have used the Henderson equation of estate 
\begin{equation}
H(\rho)=2\rho \frac{1-7\rho/16}{(1-\rho)^2}\Rightarrow \chi(\rho)=\frac{1-7\rho/16}{(1-\rho)^2}
\end{equation}
 that it known to be accurate up to a $1\%$ at the intermediate range of densities \cite{Henderson}. 
 
\begin{figure}[h!]
\begin{center}
\includegraphics[height=7cm,clip]{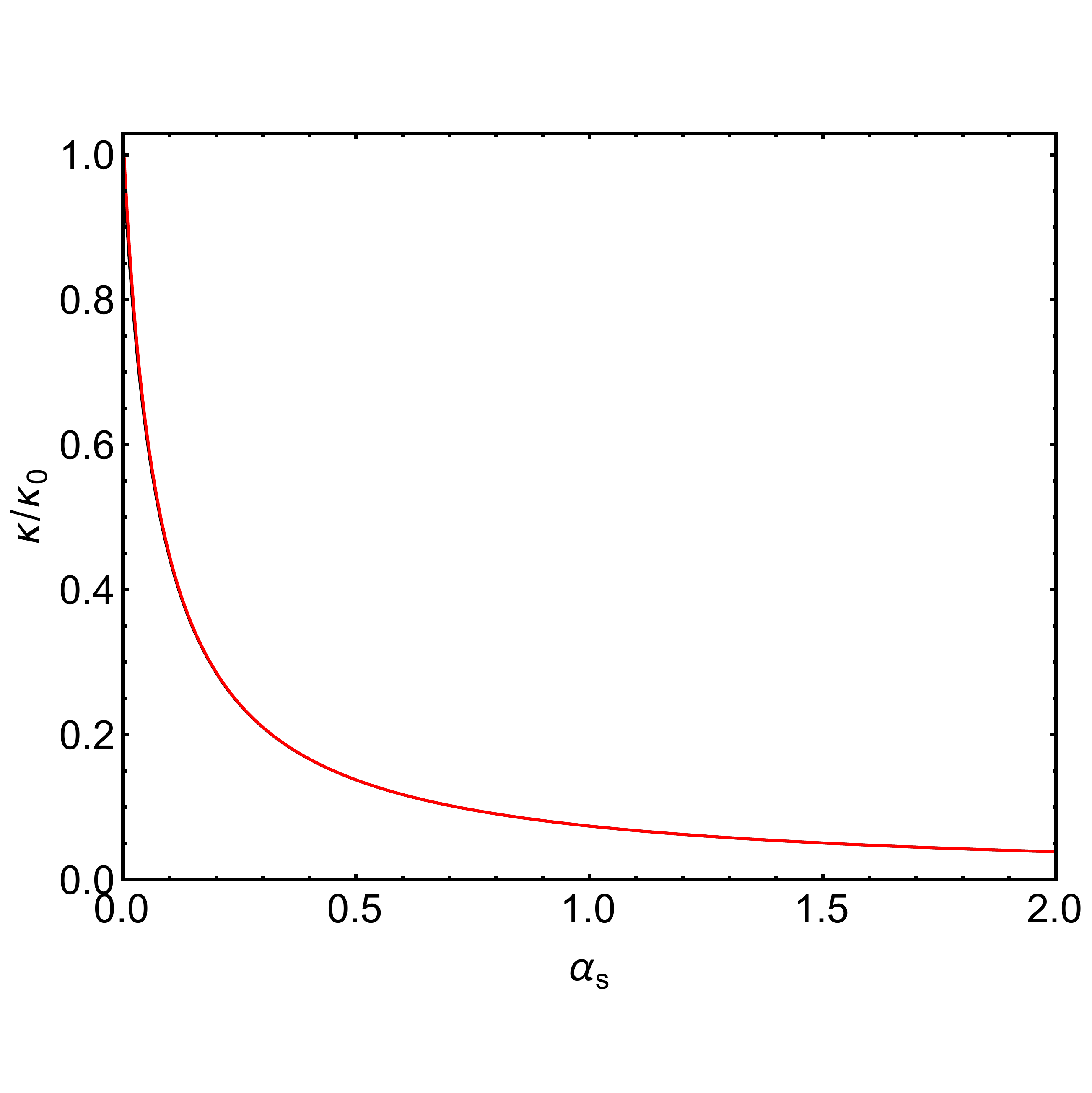}  %heat_conductivity.nb
\includegraphics[height=7cm,clip]{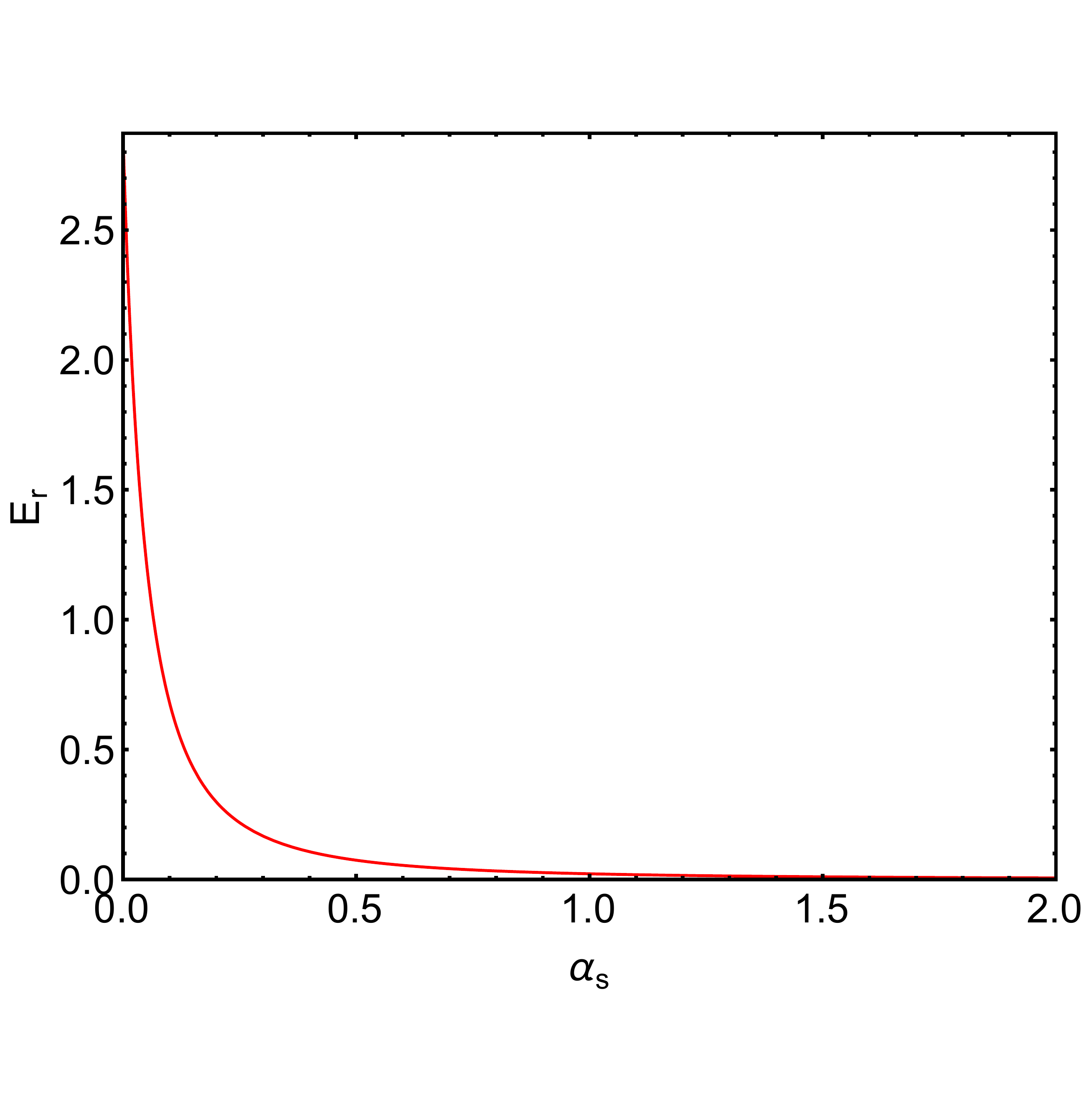}      
\end{center}
\kern -1cm
\caption{Left: $\kappa_{BGK}/\kappa_0$ (black line) and $\kappa_B/\kappa_0$ (red line) versus $\alpha_s$.  Right: Relative error: $E_r=100(1-\kappa_{BGK}/\kappa_{B})$ versus $\alpha_s$ \label{heat1}}
\end{figure}

 We can also study the heat conductivity. For the BGK case we find:
 \begin{equation}
\kappa_{BGK}=\kappa_0\frac{1}{1+4 \pi  \alpha _s}
\end{equation}
with $\kappa_0=2T^{1/2}/b\pi^{1/2}$. That is, it has no density dependence. Similarly, the Boltzmann approach has a more involved $\alpha_s$ dependence but it lacks a density dependence:
\begin{equation}
\kappa_B=\kappa_0\frac{8053063680 \tilde{\alpha }^3+12954255360 \tilde{\alpha }^2+6551619492 \tilde{\alpha }+1038480183}{32212254720 \tilde{\alpha }^4+59870085120 \tilde{\alpha }^3+39075815056 \tilde{\alpha }^2+10601599676 \tilde{\alpha }+1008650032}\nonumber
\end{equation} 
where $\tilde\alpha=\pi\alpha_s$. $\kappa_{BGK}$ and $\kappa_B$ look very different but they are not so.  We plot in figure \ref{heat1} the behavior of $\kappa_{BGK}/\kappa_0$ and $\kappa_{B}/\kappa_0$ as a function of $\alpha_s$. We do not see there any clear difference. Their relative difference $E_r$ is at most of $3\%$ and it diminishes to zero for large $\alpha_s$ values. Again, BGK seems to be a very good approximation of the Boltzmann kernel.
\begin{figure}[h!]
\begin{center}
\includegraphics[height=7cm,clip]{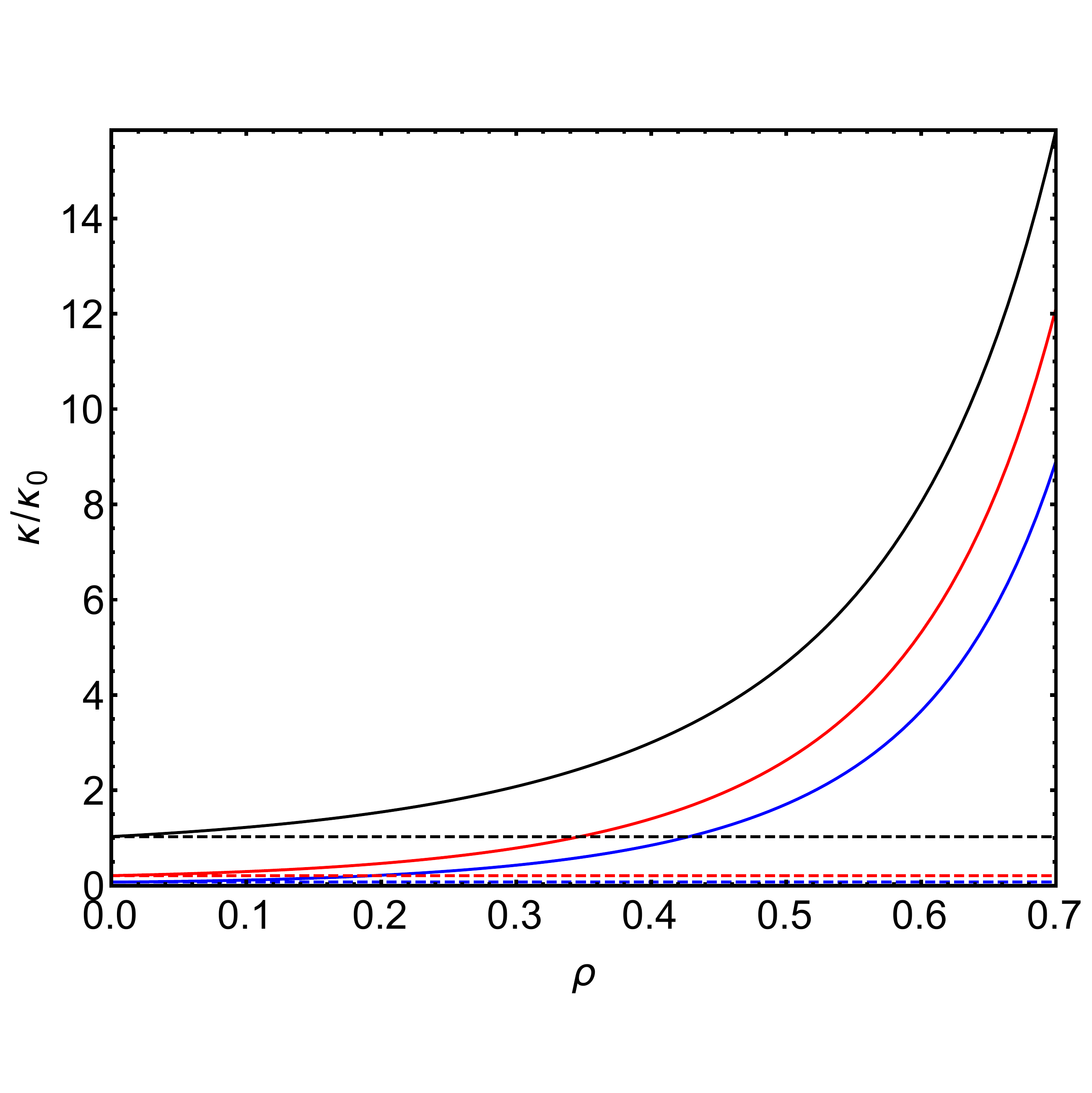}  %heat_conductivity.nb
\end{center}
\kern -1cm
\caption{Heat conductivity for the Enskog equation ($\tilde N=5$) vs areal density for different values of $\alpha_s$: $\alpha_s=0$, $0.3$ and $1$ (black, red and blue curves respectively). The dotted curves are the heat conductivity at the Boltzmann approach for the same values of $\alpha_s$ accordingly with their colors. \label{heat2}}
\end{figure}
The Enskog case has a density structure that depends on $\alpha_s$ as we show in figure \ref{heat2} for different values of $\alpha_s$ where we have used again the Henderson equation of state. For low densities the Chapman-Enskog approach and the Boltzmann one coincide. Finally let us write down explicitly the $\tilde N=2$ case
 that it is the one shown in most books for $\alpha_s=0$ (see for instance \cite{Gass}):
 \begin{equation}
\kappa^{(2)}=\kappa_0 \frac{1}{\chi+4\pi\alpha_s}\biggl[1+(3+16\rho\alpha_s)\rho\chi
+\frac{16+9\pi}{4\pi}\rho^2\chi^2
\biggr]
\end{equation}

Once we know the transport coefficients of our system at the diffusive scale we are ready to apply the Macrosocopic Fluctation Theory to it. One can show (see ref. \cite{Fox}) that it is enough to consider a local white noise matrix whose intensity is defined by the Onsager coefficients. In concrete terms the fluctuating equations are
\begin{equation}
\partial_\tau \eta_{\alpha}+\sum_i\partial_i\left[J_{\alpha ,i}+\Theta_{\alpha ,i} \right]=0
\end{equation}
where $\alpha=d,h$ and $\Theta(x,\tau)$ is a gaussian white noise field with covariances:
\begin{equation}
\langle\Theta_{\alpha ,i}(x,\tau)\Theta_{\beta ,j}(x',\tau') \rangle=2L_{\alpha\beta}\delta_{i,j}\delta(x-x')\delta(\tau-\tau')
\end{equation}
with $L_{\alpha\beta}$ being the Onsager's coefficients computed in each case.

\section{Acknowledgements}
We thank H. Spohn, C. Bernardin and specially R. Esposito and D. Gabrielli for very helpful correspondences. This work was supported in part by AFOSR [grant FA-9550-16-1-0037].
 PLG was supported also by the Spanish governement project FIS2013-43201P. We thank the IAS System Biology divison for its hospitality during part of this work.

\section*{Appendix I}

In this Appendix we study the behavior under scaling of equation:
\begin{equation}
\partial_t f+v\nabla f=\alpha\int\, dn \vert v''\cdot n\vert^s\left[f(x,v'',\tau)-f(x,v,\tau) \right]\label{a1}
\end{equation}
It is convenient to work with $v$  in polar coordinates $\tilde f(r,w,\theta,t)$ where $v=w(\cos\theta,\sin\theta)$, $w\in[0,\infty]$ and $\theta\in[0,2\pi]$. The corresponding equation for $\tilde f(r,w,\theta,t)$ is:
\begin{eqnarray}
&&\partial_t\tilde f(r,w,\theta,t)+w\left(\cos\theta\,\partial_1+\sin\theta\,\partial_2\right)\tilde  f(r,w,\theta,t)\nonumber\\
&=&2\alpha w^s\int_0^{\pi/2}d\phi\,\cos(\phi-\theta)^s\biggl[ \tilde f(r,w,\theta+2\phi+\pi,t)+\tilde f(r,w,\theta-2\phi+\pi,t)\nonumber\\
&&-2\tilde f(r,w,\theta,t)\biggr]\label{a2}
\end{eqnarray}
where $\partial_i=\partial/\partial r_i$. It is convenient now to introduce the Fourier transform of the angle coordinate $\theta$:
\begin{equation}
\hat f(r,w,l,t)=\int_{0}^{2\pi}\frac{d\theta}{2\pi}e^{-il\theta}\tilde f(r,w,\theta,t)
\end{equation}
with $l\in Z$.
 Its evolution equation is, after using eq.(\ref{a2}) ,
\begin{eqnarray}
\partial_t\hat f(r,w,l,t)&+&\frac{w}{2}\partial_1\left[\hat f(r,w,l-1,t)+\hat f(r,w,l+1,t)\right]+\frac{w}{2i}\partial_2\left[\hat f(r,w,l-1,t)-\hat f(r,w,l+1,t)\right]\nonumber\\
&=&-2\alpha w^s A(s,l) \hat f(r,w,l,t)\label{a3}
\end{eqnarray}
with
\begin{equation}
A(0,l)=\pi (1-\delta_{l,0})\quad ,\quad A(1,l)=\frac{8l^2}{4l^2-1}
\end{equation}
We now introduce the diffusive scaling: $\tau=\epsilon^2t$ and $x=\epsilon r$. We observe that $l=0$ and $l\neq 0$ have different behavior with $\epsilon$. For $l=0$ we get:
\begin{eqnarray}
&&\epsilon^2\partial_{\tau}\hat g(x,w,0,\tau)+\epsilon\frac{w}{2}\partial_1\left[\hat g(x,w,-1,\tau)+\hat g(x,w,1,\tau)\right]\nonumber\\
&&+\epsilon\frac{w}{2i}\partial_2\left[\hat g(x,w,-1,\tau)-\hat g(x,w,1,\tau)\right]=0
\end{eqnarray} 
and for $l\neq 0$:
\begin{eqnarray}
\hat g(x,w,l,\tau)&=&-\epsilon\frac{w^{1-s}}{2\alpha A(s,l)}\biggl[\frac{1}{2}\partial_1\left[\hat g(x,w,l-1,\tau)+\hat g(x,w,l+1,\tau)\right]\nonumber\\
&+&\frac{1}{2i}\partial_2\left[\hat g(x,w,l-1,\tau)-\hat g(x,w,l+1,\tau)\right]\biggr]+O(\epsilon^2)
\end{eqnarray}
where $\hat g(x,w,l,\tau)=\hat f(\epsilon^{-1}x,w,l,\epsilon^{-2}\tau)$. We observe that $\hat g(x,w,l,\tau)$ is of order $\epsilon$ when $l\neq 0$. Therefore, at the limit $\epsilon\rightarrow 0$ we can get a closed equation for $\hat g(x,w,0,\tau)$:
\begin{equation}
\partial_{\tau}\hat g(x,w,0,\tau)=D(w,\alpha,s)\Delta \hat g(x,w,0,\tau)
\end{equation}
where
\begin{equation}
D(w,\alpha,s)=\frac{w^{2-s}}{4\alpha A(s,1)}
\end{equation}
is the diffusion constant for a given particle velocity $w$. Observe that if we choose any typical particle at equilibrium $w\simeq T^{1/2}$ and this mechanism will diffuse the particle with a diffusion constant $D\simeq T^{1-s/2}$. If $s=0$ the diffusion is Brownian-like ($D\simeq T$) and $D\simeq T^{1/2}$ when $s=1$. 

\section*{Appendix II}

The BGK approximation when $\alpha=0$ and in the diffusive scaling is computed in this appendix. We take $Q(g)=Q^{BGK}$ given by eq. (\ref{BGK7}) that it has four collision invariants: $1$, $v_1$, $v_2$ and $v^2$. Assuming that $g=g_0+\epsilon g_1+\ldots$ the equations (\ref{e0}) and (\ref{e1}) read in this case:
\begin{eqnarray}
Q(g_0)&=&0\nonumber\\
v\cdot\nabla g_0&=&-\nu g_1\nonumber
\end{eqnarray}
therefore $g_0(x,v,\tau)=M(v;n(x,\tau),u(x,\tau),T(x,\tau))$ and 
\begin{equation}
g_1(x,v,\tau)=-\frac{g_0(x,v,\tau)}{v}\sum_i v_i\biggl[\frac{\nabla_i n}{n}-\frac{\nabla_i T}{T}+\frac{\nabla_i T}{2T^2}(v-u)^2+\frac{1}{T}\sum_{j}(v_j-u_j)\nabla_iu_j)
\biggr]
\end{equation}
and we can obtain the evolution equations associated with the collision invariants by using eq. (\ref{inv}):
\begin{eqnarray}
\partial_\tau n(x,\tau)+\nabla J_n(x,\tau)&=&0\nonumber\\
\partial_\tau \left[n(x,\tau)u_l(x,\tau)\right]+\nabla J_{u}^{(l)}(x,\tau)&=&0\nonumber\\
\partial_\tau \left[n(x,\tau)T(x,\tau)+\frac{1}{2}u(x,\tau)^2\right]+\nabla J_h(x,\tau)&=&0\nonumber
\end{eqnarray}
where
\begin{eqnarray}
J_{n,k}&=&-\frac{1}{\nu}\sum_i\nabla_i\left[nT\delta_{i,k}+nu_iu_k \right]\nonumber\\
J_{u,k}^{(l)}&=&-\frac{1}{\nu}\sum_i\nabla_i\left[nT(\delta_{l,k}u_i+\delta_{l,i}u_k+\delta_{k,i}u_l)+nu_iu_ku_l \right]\nonumber\\
J_{h,k}&=&-\frac{1}{\nu}\sum_i\nabla_i\left[nT(4T+u^2)\delta_{k,i}+(6T+u^2)nu_iu_k \right]\nonumber
\end{eqnarray}
These currents may be written as combinations of thermodinamic forces and Onsager's coefficients. We first use a more compact notation. We define $\eta_0=n$, $\eta_1=nu_1$, $\eta_2=nu_2$ and $\eta_3=nT+u^2/2$. The kinetic equations are then
\begin{equation}
\partial_\tau\eta_\gamma+\nabla J_\gamma=0
\end{equation}
therefore
\begin{equation}
J_{\gamma,k}=\sum_{\beta,i}L_{\gamma k,\beta i}X_{\beta i}
\end{equation}
where $X_{\beta,i}=\nabla_i y_\beta$ are the thermodynamic forces  with 
\begin{equation} 
y_0=-\frac{\mu}{T}+\frac{u^2}{2T}\quad,\quad y_i=-\frac{u_i}{T}\quad (i=1,2)\quad,\quad y_3=\frac{1}{T}
\end{equation}
(see ref. \cite{sm}). And $L$-s are the Onsager's coefficients that in our case are:
\begin{eqnarray}
L_{0k,0i}&=&\frac{n}{\nu}(T\delta_{k,i}+u_iu_k)\nonumber\\
L_{0k,ji}&=&L_{ji,0k}=\frac{n}{\nu}(T(\delta_{k,i}u_j+\delta_{k,j}u_i+\delta_{i,j}u_k)+u_iu_ju_k)\quad (j=1,2)\nonumber\\
L_{0k,3i}&=&L_{3i,0k}=\frac{n}{2\nu}(T(4T+u^2)\delta_{k,i}+(6T+u^2)u_iu_k)\nonumber\\
L_{lk,ji}&=&\frac{n}{\nu}(T^2(\delta_{j,i}\delta_{k,l}+\delta_{j,k}\delta_{l,i}+\delta_{j,l}\delta_{k,i})+T(\delta_{l,k}u_iu_j+\delta_{l,i}u_ju_k+\delta_{k,i}u_ju_l\nonumber \\
&+&\delta_{j,i}u_lu_k+\delta_{j,k}u_iu_l+\delta_{j,l}u_iu_k )+u_iu_ju_ku_l)\quad (l,j=1,2)\nonumber\\
L_{lk,3i}&=&L_{3i,lk}=\frac{n}{2\nu}(T(6T+u^2)(\delta_{l,k}u_i+\delta_{l,i}u_k+\delta_{k,i}u_l)+(8T+u^2)u_iu_ku_l)\quad (l=1,2)\nonumber\\
L_{3k,3i}&=&\frac{n}{4\nu}(T(24T^2+16Tu^2+(u^2)^2)\delta_{k,i}+(6T+u^2)(8T+u^2)u_iu_k)
\end{eqnarray}

\section*{Appendix III}

In this appendix we obtain $g_1$ for the Boltzmann collision kernel. $g_1$ is the solution of Equation (\ref{e1}) that reads
\begin{eqnarray}
&&v\cdot\left(\frac{\nabla n}{n}+\frac{\nabla T}{T}\left(\frac{v^2}{2T}-1\right)\right)=\nonumber\\
&&\frac{b}{2}\int d\hat n\int dv_2\vert (v-v_2)\cdot n \vert g_0(v_2)\left[\Phi(v')+\Phi(v_2')-\Phi(v)-\Phi(v_2) \right]\nonumber\\
&+&\alpha\int d\hat n \left[\Phi(v'')-\Phi(v) \right] \label{Q1B}
\end{eqnarray}
where we have set
\begin{equation}
g_1(x,v,\tau)=g_0(x,v,\tau)\Phi(x,v,\tau)  \label{Phi00}
\end{equation}
and we have simplified the notation: the fields $n$ and $T$ depend on $(x,\tau)$, all the other functions depend on $(x,v,\tau)$ and we only explicitly write the arguments that change and/or when it is needed to stress some fact.

We  write the unknown function $\Phi$ in the form:
\begin{equation}
\Phi=v\cdot\left[A^{(1)}(w)\frac{\nabla n}{n}+A^{(2)}(w)\frac{\nabla T}{T}\right]\label{Phi0}
\end{equation}
where $w$, $w'$, $w_2$ and $w_2'$ are the modulus of the vectors $v$, $v'$, $v_2$ and $v_2'$ respectively. Then we substitute $\Phi$ into eq. (\ref{Q1B}) and we consider  $\nabla n$ and $\nabla T$ as independent variables. Therefore we  can identify each gradient coefficient at both sides of the equation and consequently we get one equation associated to each of the gradients:
\begin{eqnarray}
G^{(i)}(w)v&=&\frac{b}{2}\int d\hat n\int dv_2\vert (v-v_2)\cdot \hat n \vert g_0(v_2)\left[A^{(i)}(w')v'+A^{(i)}(w_2')v_2'-A^{(i)}(w)v-A^{(i)}(w_2)v_2 \right]\nonumber\\
&-&2\pi\alpha A^{(i)}(w)v \quad \quad i=1,2\label{qq1}
\end{eqnarray}
where $G^{(1)}(w)=1$ and $G^{(2)}(w)=w^2/2T-1$. In order to solve these equations it is convenient to build a way to do explicitly the integrals. Therefore we expand the unknown functions $A^{(i)}$ with respect to an orthogonal polynomial base. We choose the associated Laguerre's of order $1$ \cite{Gass}:
\begin{equation}
A^{(i)}(w)=\sum_{p=0}^{\infty} a_p^{(i)}\mathcal{L}_p^{1}(\eta)\quad\quad \eta=\frac{w^2}{2T}\label{As}
\end{equation}
These set of polynomials have the form:
\begin{equation}
\mathcal{L}_p^1(\eta)=\sum_{k=1}^p (-\eta)^k \frac{(p+1)!}{(k+1)!k!(p-k)!}
\end{equation}
and they have the property:
\begin{equation}
\int_{0}^{\infty} dz e^{-z}\mathcal{L}_p^1(z)\mathcal{L}_q^1(z)z=(p+1)\delta_{p,q}
\end{equation}
Next we substitute the expanded $A^{(i)}$ functions (\ref{As}) into eq. (\ref{qq1}), we multiply both sides of the equation by $v$, by the distribution $m(v;0,T)=(2\pi T)^{-1}\exp\left[-v^2/2T\right]$ and we integrate over $v$. Thus, equation (\ref{qq1}) can be written as:
\begin{equation}
\gamma_p^{(i)}=\sum_{q=0}^{\infty}F_{pq}a_q^{(i)}\label{qq2}
\end{equation}
where
\begin{equation}
\gamma_p^{(i)}=\int dv v^2 m(v;0,T) \mathcal{L}_p^1(\eta)G^{(i)}(w)\label{alfadef}
\end{equation}
and
\begin{eqnarray}
F_{pq}&=&\int dv\, m(v;0,T)\mathcal{L}_p^1(\eta)\biggl[\frac{b}{2}n\int d\hat n\int dv_2 \vert(v-v_2)\cdot\hat n \vert m(v_2;0,T)\biggl[\mathcal{L}_q^1(\eta')v'\cdot v\nonumber\\
&+&\mathcal{L}_q^1(\eta_2')v_2'\cdot v-\mathcal{L}_q^1(\eta)v\cdot v-\mathcal{L}_q^1(\eta_2)v_2\cdot v
\biggr]-2\pi\alpha \mathcal{L}_q^1(\eta)v\cdot v
\biggr]
\end{eqnarray}
where $\eta'=v'^2/2T$, $\eta_2=v_2^2/2T$ and $\eta_2'=v_2'^2/2T$. Hence, $g_1$ is known once we obtain the $a_q^{(i)}$'s that depend on the inversion of an infinite dimensional square matrix. The solution is approximated  by truncating the infinite matrix to a $\tilde N$-dimensional one and studying its convergence when $\tilde N\rightarrow\infty$. The approximate $\tilde N$-order equation is then written as:
\begin{equation}
\gamma_p^{(i)}=\sum_{q=0}^{\tilde N}F_{pq}a_q^{(i,\tilde N)}\label{qq3}\quad p=1,\ldots,\tilde N
\end{equation}
whose solution is given by
\begin{equation}
a_q^{(i,\tilde N)}=\sum_{p=0}^{\tilde N} (F^{-1})_{qp}\gamma_p^{(i)}\label{qq4}
\end{equation}
or in matrix notation 
\begin{equation}
a^{(i,\tilde N)}=F_{\tilde N}^{-1}\gamma^{(i,\tilde N)}\label{qq5}
\end{equation}
where $\gamma^{(i,\tilde N)}$ is the $\tilde N$'th order dimensional vector built with the first $\tilde N$ coefficients of $\gamma^{(i)}$.
When equations (\ref{qq5}), (\ref{As}), (\ref{Phi0}) are substituted into eq. (\ref{Phi00}) it gives $g_1$ at the $\tilde N$-approximation. Therefore, the currents for the diffusion equations for $n$ and $T$ (\ref{dif}) are written in this case as:
\begin{eqnarray}
J_n^{(\tilde N)}&=&\frac{1}{2}\bar a_{11}^{(\tilde N)}\nabla n+\frac{n}{2T}\bar a_{12}^{(\tilde N)}\nabla T\nonumber\\
J_h^{(\tilde N)}&=&\frac{T}{2}(\bar a_{11}^{(\tilde N)}+\bar a_{12}^{(\tilde N)})\nabla n+\frac{n}{2}(\bar a_{12}^{(\tilde N)}+\bar a_{22}^{(\tilde N)})\nabla T\label{currentB}
\end{eqnarray}
where 
\begin{equation}
\bar a_{ij}^{(\tilde N)}={\gamma^{(i,\tilde N)}}^T F_{\tilde N}^{-1}\gamma^{(j,\tilde N)}\label{abar}
\end{equation}
Therefore, the diffusion coefficient and the thermal conductivity are given by
\begin{equation}
D^{(\tilde N)}=-\frac{1}{2}\bar a_{11}^{(\tilde N)}\quad,\quad \kappa^{(\tilde N)}=\frac{n}{2\bar a_{11}^{(\tilde N)}}(\bar a_{12}^{(\tilde N)2}-\bar a_{11}^{(\tilde N)}\bar a_{22}^{(\tilde N)})
\end{equation}
Finally, the Onsager's coefficients in the $\tilde N$ approximation are, 
\begin{equation}
L_{nn}^{(\tilde N)}=-\frac{n}{2}\bar a_{11}^{(\tilde N)}\, ;\, L_{nh}^{(L)}=L_{hn}^{(L)}=-\frac{nT}{2}(\bar a_{11}^{(\tilde N)}+\bar a_{12}^{(\tilde N)})\, ;\, L_{hh}^{(\tilde N)}=-\frac{nT^2}{2}(\bar a_{11}^{(\tilde N)}+2\bar a_{12}^{(\tilde N)}+\bar a_{22}^{(\tilde N)})\label{Ons2}
\end{equation}
One can get explicitely the values of the $\gamma$'s coefficients and the components of the square symmetric matrix $F_{\tilde N}$:
\begin{equation}
\gamma_q^{(1)}=2T\delta_{q,0}\quad\quad,\quad\quad \gamma_q^{(2)}=2T(\delta_{q,0}-\delta_{q,1})
\end{equation}
and
\begin{equation}
F=T\left[bn(4\pi T)^{1/2}\tilde F_C-4\pi\alpha\tilde F_D \right]
\end{equation}
with
\begin{equation}
\tilde F_C=\left(\begin{array}{cccccc}
0&0&0&0&0&\ldots\\
0&-2&1/2&1/16&1/64&\ldots\\
0&1/2&-39/8&91/64&53/256&\ldots\\
0&1/16&91/64&-4433/512&5435/2048&\ldots\\
0&1/64&53/256&5435/2048&-108335/8192&\ldots\\
\ldots&\ldots&\ldots&\ldots&\ldots&\ldots\\
\end{array}\right)
\end{equation}
and
\begin{equation}
(\tilde F_D)_{pq}=(p+1)\delta_{p,q}\quad \quad p,q=0,1,\ldots
\end{equation}
We observe that for any $\tilde N$ value we can compute the coefficients of the inverse matrix $F$:
\begin{equation}
(F_{\tilde N}^{-1})_{00}=-\frac{1}{4\pi\alpha T}\quad,  (F_{\tilde N}^{-1})_{01}=0
\end{equation}
which implies
\begin{equation}
\bar a_{11}=\bar a_{12}=-\frac{T}{\pi\alpha}
\end{equation}

\section*{Appendix IV}

 In this appendix we compute $g_1$ for the Enskog collision kernel. We solve the equation:
\begin{equation}
v\cdot\nabla g_0=Q_{C,0;1}^{E}(g_0,g_1)+Q_{D;1}(g_1)+Q_{C,1;0}^{E}(g_0)\label{e34}
\end{equation}
with functions $Q_{C,0;1}^E$, $Q_{D;1}(g_1)$ and $Q_{C,1;0}^{E}(g_0)$ defined in eqs. (\ref{exp1}) and (\ref{exp2}).
Observe that the two first terms on the right hand side  are the same as we had at order $\epsilon$ for the Boltzmann equation kernel above, eq.  (\ref{Q1B}), with just the inclusion of $\chi$ in front of $Q_C$ there. Moreover, the last term in eq. (\ref{e34}), $Q_{C,1;0}^{E}$, just depends on $g_0$, that is, only depends on $n$ and $T$ fields and we can put them together with ones similar on the left hand side of eq. (\ref{e34}). In fact, one can show that
\begin{equation}
Q_{C,1;0}^{E}=-\pi \frac{b^2}{2}n g_0 \chi v\cdot\left[\frac{\nabla\chi}{\chi}+2\frac{\nabla n}{n}+\frac{\nabla T}{2T}\left(\frac{3v^2}{4T}-1\right)\right]
\end{equation}
 Therefore, if we assume $g_1=g_0\Phi$, we can write a set of very similar equations as we did for the Boltzmann case at order $\epsilon$:
\begin{eqnarray}
&&v\cdot\left(f_1(n)\nabla n+\frac{\nabla T}{T}\left(f_2(n)\frac{v^2}{2T}-f_3(n)\right)\right)=\nonumber\\
&&\frac{b}{2}\int d\hat n\int dv_2\vert (v-v_2)\cdot n \vert g_0(v_2)\left[\Phi(v')+\Phi(v_2')-\Phi(v)-\Phi(v_2) \right]\nonumber\\
&+&\alpha\int d\hat n \left[\Phi(v'')-\Phi(v) \right] \label{CE3}
\end{eqnarray}
where
\begin{eqnarray}
f_1(n)&=&\frac{1}{n}\left[\frac{1}{\chi}+\pi b^2 n+\pi b^2 n^2\frac{\chi'}{2\chi} \right]\nonumber\\
f_2(n)&=&\frac{1}{\chi}+\frac{3}{8}\pi b^2n\nonumber\\
f_3(n)&=&\frac{1}{\chi}+\frac{1}{4}\pi b^2n
\end{eqnarray}
where we have changed the value of $\alpha$ from the Boltzmann kernel case: $\alpha\rightarrow\alpha\chi$ and we have assumed that $\chi$ depends on $r$ only through the density field, $\chi=\chi(n)$ and therefore $\chi'=d\chi/dn$. 

We now suppose that
\begin{equation}
\Phi=v\cdot\left[B^{(1)}(w)f_1(n)\nabla n+B^{(2)}(w)f_2(n)\frac{\nabla T}{T} \right]
\end{equation}
and substituting this expression into eq. (\ref{CE3}) and identifying gradients as in the Boltzmann kernel case, we get the same equations (\ref{qq1}) with the changes: 
\begin{equation}
G^{(i)}\rightarrow H^{(i)}\,;\,A^{(i)}\rightarrow B^{(i)} 
\end{equation}
where $H^{(1)}(w)=1$ and $H^{(2)}(w)=w^2/2T-f_3(n)/f_2(n)$. At this point we follow the same path as in the Boltzmann kernel case to solve the equations only taking into account the functional differences pointed out. We decompose $B^{(i)}$ with respect of the associated Laguerre polynomials:
\begin{equation}
B^{(i)}(w)=\sum_{p=0}^{\infty}b_p^{(i)}\mathcal{L}_p^1(\eta)\quad \eta=\frac{w^2}{2T}
\end{equation}
the equation (\ref{CE3}) is then written as
\begin{equation}
\beta_p^{(i)}=\sum_{q=0}^{\infty}F_{pq}b_q^{(i)}
\end{equation}
with
\begin{equation}
\beta_q^{(1)}=2T\delta_{q,0}\quad ,\quad \beta_q^{(2)}=2T\left(\left(2-\frac{f_3(n)}{f_2(n)}\right)\delta_{p,0}-2\delta_{p,1}\right)
\end{equation}
Observe that the coefficients of the matrix $F$, $F_{pq}$, are the same as in the Boltzmann kernel case. The $\tilde N$'th truncated solution is given by
\begin{equation}
b_q^{(i,\tilde N)}=\sum_{p=0}^{\tilde N}(F^{-1})_{pq}\beta_{p}^{(i)}
\end{equation}

We can get the values of $\bar b_{ij}^{(\tilde N)}$ from the already computed ones, $\bar a_{ij}^{(\tilde N)}$, from the Boltzmann kernel case. We see that $\beta_q^{(1)}=\alpha_q^{(1)}$ and $\beta_q^{(2)}=\alpha_q^{(2)}+2T(1-D)\delta_{q,0}$. Therefore, it is a simple matter of algebra to find:
\begin{eqnarray}
\bar b_{11}&=&\bar a_{11}=-\frac{T}{\pi\alpha}\nonumber\\
\bar b_{12}&=&\bar a_{12}+\left(1-\frac{f_3(n)}{f_2(n)}\right)\bar a_{11}=-\frac{T}{\pi\alpha}\left(2-\frac{f_3(n)}{f_2(n)}\right)\nonumber\\
\bar b_{22}^{(\tilde N)}&=&\bar a_{22}^{(\tilde N)}+2\left(1-\frac{f_3(n)}{f_2(n)}\right)\bar a_{12}+\left(1-\frac{f_3(n)}{f_2(n)}\right)^2\bar a_{11}\nonumber \\
&=&\bar a_{22}^{(\tilde N)}+\frac{T}{\pi\alpha}\frac{f_3(n)}{f_2(n)}\left(2-\frac{f_3(n)}{f_2(n)}\right)\label{bb}
\end{eqnarray}
where $\bar b_{11}$ and $\bar b_{12}$ do not depend on the $\tilde N$ as in the Boltzmann case.

\end{document}